# A Study of the Time Variability and Line Profile Variations of κ Dra

S. M. Saad[1,2], M. I. Nouh[1], A. Shokry[1,2] and I. Zead[1,2]

[1]*National Research Institute of Astronomy and Geophysics, 11421 Helwan, Cairo, Egypt*
[2]*Kottamia Center of Scientific Excellence for Astronomy and Space Research KCScE STDF 5217.*

**Abstract:** In this study, we present a spectroscopic analysis of the bright Be star κ Dra. Two independent sets of radial velocity velocities (RV) measurements were obtained by direct measurement and using a line profile disentangling technique. Using a combination of the solutions found by the codes **FOTEL** and **KOREL**, we revisited the binary nature of κ Dra and derived improved orbital elements. From the RVs of the Balmer lines and also from some strong metallic lines we found that all RVs variations are phase-locked with the orbital period. V/R variations were obtained for $H_\alpha, H_\beta, H_\gamma$ and some other photospheric lines and they are found to be phase-locked with the orbital motion. A moving absorption bump superimposed over the $H_\alpha$ and $H_\beta$ emission line profiles was detected. The orbital solutions for κ Dra were derived assuming a circular orbit with a period $P = 61^d.5549$ and K = 6.81 $Kms^{-1}$. The question of the line profile variability was discussed. We attempted to search for absorption or emission lines of the unresolved secondary component, we failed to find them.

## 1. Introduction:

The system κ Dra (5 Dra, HD 109387, HR 4787, BD+70 703, MWC 222, HIP 61281, Gaia DR2 1683102889080253312) is a bright variable (*m*v = 3.75–3.95), Be star known as a single-lined spectroscopic binary. The binary nature of κ Dra has been reported a long time ago by Hill (1926) and later by Miczaika (1950) based on radial velocity (RV) variations. The first orbital solution was obtained by Juza et al. (1991). Based on an analysis of RV measurements, they derived an orbital period of $61^d.5549$ and found phase-locked V/R variations of the double peak Balmer emission lines. In addition, information about eccentricity, semi-amplitude, and mass functions was obtained. However, some mystery about the candidate of the secondary star remains. Saad et al. (2005) suggest that κ Dra is a circular binary with H α and H β emission locked with the orbital period. Juza et al. (1994) show that the maximum of the emission strength is preceded by the long-term cycle maximum brightness and coincides with the maximum of the continuum polarization (Arsenijevic et al. 1994). Hirata (1995) investigated the brightening of the object and concluded that the extended photosphere may be modeled by rotating model photospheres. Hill et al. (1991) investigated available spectra and showed that a period of $0^d545$ interprets the rapid variation better than the period of $0^{.d}89$ period computed by Juza et al. (1991). Saad et al. (2004) derived the stellar mass $M = 4.8 \pm 0.8\ M_\odot$ and radius $R = 6.4 \pm 0.5\ R_\odot$ for the primary.

In the present paper, we analyze the spectroscopic data of the bright Be star *κ* Dra obtained in the period (1994-2003). We use KOREL and FOTEL codes to compute the orbit of the system. Time variability, line profile variations, and the V/R variations are analyzed.

## 2. Observations and data reduction

Our data consist of three sets of electronic spectra obtained between June 1992 and April 2003 in both coudé and Cassegrain foci of the Ondřejov Perek 2-m telescope between HJD 48813.4316 and 52754.8733.

- The main set consists of 102 spectra, which mainly cover the $H_\alpha$ region in a spectral range 6300-6700Å and were obtained using the Reticon detector in the coudé focus.
- The second set is a series of simultaneous red 5850-7850Å and blue 3450-5650Å spectra secured with the HEROS echelle spectrograph in the Cassegrain focus.
- Several CCD spectrograms in both red 6300-6740Å and blue 4753-5005Å regions were obtained in the coudé focus.

The main features of the spectrograms and a description of used detectors are summarized in Saad et al. (2004), where the details about the main reduction procedures are also discussed. Table 1 lists the journal of the spectroscopic observations of the system.

Table 1. The observational journal of κ Dra at the Ondřejov Observatory (taken from Saad et al. 2004)

| Epoch (HJD-2400000) | No. of Spectra | Spectrograph | Detector | Resolving Power | Spectral Range [Å] |
|---|---|---|---|---|---|
| 48000- 51714 | 93 | coudé slit | Reticon RL-1872F/30 | 10 000 | 6300 – 6740 |
| 49021-49026 | 3 | coudé slit | Reticon RL-1872F/30 | 10 000 | 4310 – 4520 |
| 49079-49116 | 6 | coudé slit | Reticon RL-1872F/30 | 10 000 | 4750 – 4960 |
| 52321-52321 | 1 | coudé slit | CCD SITe005 800×2000 | 10 000 | 4300 – 4554 |
| 52323-52323 | 2 | coudé slit | CCD SITe005 800×2000 | 10 000 | 6256 – 6769 |
| 51900-52727 | 30 | HEROS | CCD EEV 2000×800 | 20 000 | 3450 – 5650 |
| | | | CCD EEV 1152×770 | 20 000 | 5850 – 8620 |
| 52742-52754 | 6 | coudé slit | CCD SITe005 800×2000 | 10 000 | 4753 – 5005 |
| 52734-52754 | 22 | coudé slit | CCD SITe005 800×2000 | 10 000 | 6256 – 6769 |

## 2.1 Radial Velocity Measurements

The RV measurements were performed interactively using the computer program **SPEFO** (developed by the late Dr. Jiří Horn) by matching the original and mirrored line profiles. The RV analysis is based mainly on the $H_\alpha$ line, where RVs are measured in several parts of the line (namely in steep parts of the line wings, the central absorption, the absorption bump, and the violet and red peaks displacements) and in strong lines of other elements in its vicinity such as He I 6678Å and Si II 6347, 6371Å. In addition, for HEROS (i.e. the wide range) spectra, RVs for Balmer lines up to H7, and other metallic lines such as Mg II 4481 Å and He I 4471Å were determined. In the latter cases, RVs for these lines were measured for the outermost parts of absorption wings, neglecting the details in the line core. Measurement of RVs in $\kappa$ Dra was not an easy task due to the low amplitude of the variations (only a few km/s) and the line-profile distortion.

The measured RVs for different lines are listed in Table 2, where the file ID, Heliocentric Julian Date of observations (HJD), and the RVs in km/s for each line are listed.

## 2.2 V/R measurements

As we have already discussed in detail in Saad et al. (2004), the optical and near infrared spectra of $\kappa$ Dra in the region 3450-7850Å are characterized by the presence of emission in lower Balmer lines ($H_\alpha$, $H_\beta$, $H_\gamma$), in some Fe II lines, and in the oxygen triplet line OI 7772, 7774, 7775Å. All the above mentioned lines display double-peaked profiles. The V/R ratio is obtained for these lines whenever it is measurable. The violet (V) and red (R) peak intensities of $H_\alpha$ are variable on a long-timescale (see Fig.4 a, b in Saad et al. (2004). The intensity ratio V/R (concerning the normalized continuum) was found to be variable with time, however, the first inspection of its time plot reveals that it varies independently of other line parameters (strength and intensity), which show cyclic long-term variations on a time scale of decades. In some phases, it was quite difficult to identify the double-peak structure of the $H_\alpha$ line or one of its peaks. For both $H_\beta$ and $H_\gamma$ lines, V/R variations were obtained over the last three years of the observations, since older Reticon spectra do not cover corresponding spectral regions. Their V/R variations generally behave similarly to those of $H_\alpha$. V/R variations of $H_\beta$ show stronger R phases than V for the same epochs and its value is longer below the unity (V/R<1), in addition to its lower amplitude scale. Due to the weak emission feature of the $H_\gamma$ line, V/R variations are not significant and the values are distributed almost around the unity. Fig.1 illustrates temporal V/R variations for the $H_\alpha$, $H_\beta$, and $H_\gamma$ lines. Moreover, V/R variations were found for some other photospheric lines affected by emission, namely for the iron lines Fe II 7462, 7712 Å, and for the triplet O I 7772-5 Å. Their relative V and R intensities are very weak, 4-6 % of the continuum level for the Fe II 7712 Å and O I 7772-5 Å lines, while 1-3 % for the Fe II 7462 Å line. During our observations, the Fe II 7462 Å line has phases dominated by relatively strong R peaks, in contradiction to the FeII 7712 Å line, where the ratio V/R is almost around unity. For the O I 7772-5 Å line, V/R displays consequent slight changes from values >1 to <1.

Table 2: Radial velocity measurements of the Hα region

| File ID | HJD -2400000 | $H_\alpha$ | | | | | Si ll 6347 | Si ll 6371 | He I 6678 |
|---|---|---|---|---|---|---|---|---|---|
| | | Wing | Abs. Cent. | V-Peak | R-Peak | Abs. Bump | | | |
| 311 | 48813.4316 | 1.5 | 36.6 | -50.6 | 71.7 | -120.2 | 14.18 | 10.85 | -7.88 |
| 685 | 48883.5011 | -2.3 | 7 | -39.8 | 44.5 | -101.8 | -2.64 | -5.89 | -18.3 |
| 743 | 48893.2295 | -2.8 | -12.7 | -53.7 | 40 | / | -10.48 | -11.43 | -14.81 |
| 939 | 49018.5005 | -0.46 | -13.93 | -61.96 | 48.15 | / | / | / | / |
| 940 | 49018.5748 | -0.5 | -16.6 | -64.3 | 48.7 | / | -12.82 | -7.29 | -2.39 |
| 1270 | 49045.4876 | -2.59 | / | / | / | / | 3.04 | -0.1 | 1.59 |
| 1303 | 49066.4515 | 2.75 | / | / | / | / | 2.84 | 4.63 | 1.44 |
| 1381 | 49079.5351 | 2.5 | -14 | -61.4 | 51.7 | / | -3.5 | 1.32 | -1.28 |
| 1407 | 49080.3336 | / | / | / | / | / | -4.06 | / | -6.19 |
| 1442 | 49081.4162 | / | / | / | 45.7 | / | -8.38 | -9.24 | -10.24 |
| 1448 | 49081.4524 | -1.7 | -14 | -65.6 | 58 | / | -10.03 | -4.31 | -10.85 |
| 1465 | 49088.3751 | 1.7 | -10 | -56.9 | / | -127.7 | -8.87 | 6.36 | -0.72 |
| 1515 | 49092.4615 | 1.9 | / | / | / | / | -4.83 | -10.58 | -7.8 |
| 1565 | 49102.3385 | 7.72 | / | / | 56.5 | / | 0.73 | 2.04 | -1.81 |
| 1566 | 49102.3636 | 5 | -12.6 | -54.2 | 53.5 | / | 0.72 | 0.83 | -0.09 |
| 1568 | 49102.3819 | 3.7 | -10.9 | -57.2 | 60.83 | / | -1.53 | 5.3 | -3.55 |
| 1697 | 49116.4731 | -1.00 | 25 | -50.01 | 68.7 | -127.13 | 11.48 | -0.88 | -2.34 |
| 1740 | 49119.5658 | 0.7 | 24.2 | -54.3 | 62.3 | -119.3 | 8.98 | -1.35 | -1.24 |
| 1849 | 49133.4032 | -4.5 | 10.7 | -60.2 | / | / | -9.34 | -20.15 | -11.36 |
| 3466 | 49310.5844 | - | / | / | 57.8 | / | -1.67 | 10.93 | -5.29 |
| 3467 | 49310.6279 | -2.5 | 14.5 | -52.9 | 42.3 | -111.5 | -1.08 | -1.15 | 2.75 |
| 3601 | 49350.6331 | -4 | -9.9 | -57.9 | / | / | -9.66 | -11.72 | -10.06 |
| 3848 | 49403.6115 | / | / | / | 57.8 | / | -8.06 | 0.46 | -5.74 |
| 4243 | 49463.3294 | -7.2 | -6 | -63.5 | 58 | -129.7 | -12.23 | -16.24 | -9.07 |

| File ID | HJD -2400000 | Wing | Abs. Cent. | V-Peak | R-Peak | Abs. Bump | Si ll 6347 | Si ll 6371 | He I 6678 |
|---|---|---|---|---|---|---|---|---|---|
| 4249 | 49463.3564 | -6.5 | -6.2 | -65.4 | 58.3 | -130.4 | -9.89 | -16.32 | -4.97 |
| 4261 | 49463.4929 | -5 | -7.9 | -60 | 58.5 | -129.2 | -6.3 | -4.9 | -10.14 |
| 4274 | 49463.5259 | -6 | -3 | -62.8 | 59.4 | -128.4 | -10.95 | -11.29 | -10.16 |
| 4406 | 49466.3311 | -5.6 | -3.9 | -66 | / | -116.4 | -9.1 | -8.34 | -5.44 |
| 4407 | 49466.3459 | / | / | / | 58.6 | / | -9.71 | -7.74 | -14.08 |
| 4427 | 49466.3994 | -6.5 | -6.5 | -65.6 | 54.8 | / | -7.41 | -9.08 | -9.07 |
| 4433 | 49466.4142 | -9.1 | -10.3 | -68.3 | 57.1 | -126.9 | -14.94 | -13.45 | -13.37 |
| 4445 | 49466.5338 | -8 | -8.5 | -68.3 | 57.1 | -125.2 | -15.47 | -16.51 | -9.42 |
| 4446 | 49466.5492 | -7.3 | -7.6 | -66.5 | 60.4 | -118.7 | -14.96 | -12.3 | -15.11 |
| 4462 | 49466.6149 | -6.4 | -7 | -66.1 | 58.7 | -115.4 | -14.6 | -7.07 | -10.35 |
| 4469 | 49467.3034 | -6.9 | -10.4 | -63.2 | 56.8 | -122.31 | -5.62 | -11.49 | -8.49 |
| 4484 | 49467.3497 | -9.5 | -11.8 | -65.7 | 50 | -124.9 | -9.83 | -3.58 | -10.14 |
| 4496 | 49467.4651 | -8.6 | -11.5 | -64.8 | 60.8 | / | -10.93 | -3.49 | -11.41 |
| 4515 | 49467.5566 | -6 | -9.5 | -62.2 | 37 | / | -9.24 | -10.83 | -11.41 |
| 7348 | 49702.655 | -6.54 | -15.33 | -62 | / | / | -4.7 | -9.21 | -13.25 |
| 8124 | 49853.4086 | / | / | / | / | / | -3.01 | -0.19 | 2.38 |
| 8126 | 49853.4793 | / | / | / | 67.9 | / | -1.55 | -0.56 | 2.68 |
| 8190 | 49862.3759 | -3.6 | 25.8 | -61 | 67.4 | -115.5 | -3.78 | -1.71 | -0.87 |
| 8191 | 49862.3842 | -6.4 | 25.2 | -60.9 | 72.8 | -117.8 | 5.31 | -6.54 | 0.86 |
| 8551 | 49918.3372 | 0.2 | 37.7 | -56.1 | 73.7 | -125.8 | -2.03 | -10.58 | -2.99 |
| 8739 | 49930.3387 | -3.1 | 28 | -55.3 | 53.8 | -112.1 | 0.27 | -3.73 | -0.21 |
| 9690 | 50080.6261 | -5.4 | -10.7 | -57.5 | / | / | -7.33 | -12.9 | -5.58 |
| 9798 | 50097.6859 | -0.67 | / | / | 68.6 | / | / | / | / |
| 9862 | 50104.7034 | -1.7 | 19.4 | -57.9 | 57.1 | -112.4 | 2.15 | -6.38 | -1.17 |
| 10089 | 50140.59 | -3.2 | -6.8 | -59.5 | 58.1 | -111.6 | -7.13 | -15.41 | -9.87 |
| 10245 | 50158.5779 | -2.2 | 27.1 | -57.2 | | / | 12.9 | 2.41 | 7.12 |

Table 2: Continued

| File ID | HJD -2400000 | H$_\alpha$ | | | | | Si ll 6347 | Si ll 6371 | He I 6678 |
|---|---|---|---|---|---|---|---|---|---|
| | | Wing | Abs. Cent. | V-Peak | R-Peak | Abs. Bump | | | |
| 10268 | 50159.4882 | -1 | 24.7 | -59.5 | 61.6 | / | 7.89 | 2.99 | -0.39 |
| 10285 | 50160.4201 | -3.1 | 21.4 | -64 | 64.7 | / | -2.49 | 3.45 | 1.05 |
| 10562 | 50193.4706 | -7.4 | -25.6 | -77.1 | 52.3 | -123.3 | -18.45 | -24.42 | -11.36 |
| 10972 | 50249.4667 | -1.4 | -13.7 | -71.6 | 57.7 | / | -11.66 | -13.96 | -10.7 |
| 10995 | 50251.383 | -4.8 | -20.6 | -75 | 50.8 | / | -16.16 | 1.52 | -0.48 |
| 11509 | 50316.3418 | -3.1 | -23.6 | -68.1 | 43.1 | -117.2 | -1.38 | -21.16 | -14.34 |
| 12257 | 50448.6371 | -4.8 | -14.2 | -66.3 | 56.1 | / | -7.81 | -16.21 | -7.85 |
| 12528 | 50497.4534 | -6.0 | -17.25 | -70.5 | 47.11 | -122 | 22.2 | / | 0.33 |
| 12576 | 50506.5231 | -7 | -8.7 | -70.2 | 58.6 | / | -9.7 | -14.04 | 0.35 |
| 12619 | 50509.529 | -6.6 | -9.5 | -69.8 | 60.1 | -130.6 | -14.86 | -16.93 | -6.8 |
| 12743 | 50518.464 | -5.3 | -8.2 | -69.7 | 64.3 | / | -7.35 | -0.23 | -1.39 |
| 13044 | 50583.4009 | -1.4 | 12.1 | -64.6 | 74.1 | / | 0.93 | -0.3 | 2.42 |
| 13401 | 51081.3122 | / | / | / | / | / | / | / | / |
| 13544 | 51238.5568 | -5.45 | -15.98 | -66.92 | 56.03 | / | / | -24.21 | 1.47 |
| 13622 | 51250.4931 | -4.8 | -15.9 | / | / | / | 0.15 | -3.89 | -5.41 |
| 13741 | 51304.5682 | / | / | / | / | / | -15.44 | / | / |
| 13795 | 51316.368 | -2.75 | -10.94 | -69.49 | 65.17 | / | / | 8.01 | -7.46 |
| 13799 | 51316.4734 | -1.26 | -14.14 | -67.42 | 67.24 | / | -3.75 | / | 15.85 |
| 13842 | 51322.4657 | / | / | / | / | / | -4.56 | / | 3.11 |
| 13849 | 51323.3312 | -1.3 | 11.6 | -61 | 74.2 | / | -2.23 | -13.11 | -2.52 |
| 13851 | 51323.3469 | -0.4 | 12.5 | -61.8 | 72.2 | / | -8.57 | -17.03 | -0.39 |
| 13853 | 51323.3652 | -0.5 | 10.6 | -62 | 73.2 | / | -7.68 | -12.52 | -1.91 |
| 13855 | 51323.3851 | 0.9 | 10.3 | -60.5 | 73.5 | / | -1.38 | / | 3.72 |
| 13857 | 51323.4095 | 1.5 | 9.1 | -61.7 | 72.9 | / | -13.26 | -26.47 | -0.73 |

| File ID | HJD -2400000 | Wing | Abs. Cent. | V-Peak | R-Peak | Abs. Bump | Si ll 6347 | Si ll 6371 | He I 6678 |
|---|---|---|---|---|---|---|---|---|---|
| 13859 | 51323.4402 | -0.1 | 9.3 | -61.6 | 73.6 | / | -6.74 | -26.62 | -4.28 |
| 13868 | 51325.3803 | -2.1 | 23.6 | -58.3 | 71.6 | / | -1.27 | -11.64 | 10.26 |
| 13920 | 51328.4117 | -0.2 | 43.7 | -59.9 | 79.4 | / | -7.14 | 16.15 | 3.05 |
| 13922 | 51328.4401 | -0.5 | 44 | -61.4 | 77.9 | / | -1.92 | -5.21 | 3.3 |
| 14118 | 51378.5441 | 0.4 | 12.1 | -62.2 | 81.8 | -112.5 | -0.97 | -18.53 | -8.52 |
| 14137 | 51379.4291 | -2.5 | 14.5 | -62.2 | 77.6 | / | 7.58 | -14.94 | -3.97 |
| 14276 | 51391.4448 | -0.8 | 51.2 | -58.8 | 70 | -116.7 | 0.51 | -0.75 | -1.69 |
| 14438 | 51401.3442 | 2.5 | 60.2 | -50.5 | 84.2 | -113.8 | / | / | 9.37 |
| 14477 | 51410.3629 | -1.5 | 27.7 | -58.9 | 63.4 | -112.2 | -9.37 | -20.37 | -4.52 |
| 14787 | 51433.3458 | -5.2 | -8.1 | -59 | 45.1 | / | -26.19 | -5.62 | -9.43 |
| 15597 | 51580.6094 | -0.1 | 56.7 | -55.1 | 74.8 | / | 4.71 | -2 | 7.94 |
| 15740 | 51602.4666 | -1 | -19.2 | -61.9 | 45.8 | / | -10.27 | -23.28 | 0.77 |
| 15898 | 51643.4587 | -1.1 | 35.8 | -52 | 66.2 | / | 3.48 | / | 8.4 |
| 15973 | 51565.3361 | -2.2 | 15.4 | -53.7 | 66.9 | / | -1.3 | -14.39 | -2.8 |
| 16003 | 51661.4248 | -4 | -21.6 | -62.6 | 45.1 | / | -16.32 | -25.03 | -1.15 |
| 16079 | 51669.3847 | -3.3 | -17.3 | -67.1 | 46.5 | / | -10.09 | / | -1.61 |
| 16137 | 51678.3676 | -4.2 | -4.2 | -54.5 | 55.5 | -123.6 | -21.83 | / | 3.75 |
| 16148 | 51679.3875 | -4.1 | -4.1 | -57.3 | 62.6 | -124.6 | -5.28 | -15.89 | -0.34 |
| 16150 | 51679.4007 | -4.5 | -0.4 | -57.8 | 62.8 | -124.5 | -3.6 | / | 0.1 |
| 16153 | 51679.4206 | -4.9 | 2.1 | -56.4 | 61.8 | -122 | -9.99 | -17.56 | -4.21 |
| 16154 | 51679.4386 | -4.6 | -1.7 | -56.7 | 62.7 | -125.2 | -16.04 | -27.21 | -1.92 |
| 16167 | 51680.3669 | -8.2 | -3.8 | -57.6 | 53.6 | -129 | -1.52 | -28.23 | -1.41 |
| 16195 | 51681.4058 | -7.3 | 4.4 | -59.9 | 58.3 | -127.8 | -9.01 | -19.65 | -14.84 |
| 16505 | 51714.3661 | -4.2 | 28 | -52.2 | 67.2 | -110.1 | -11.34 | / | -3.78 |
| 17134 | 51924.5382 | -7.6 | / | -53.3 | 46.7 | -114.5 | -4.97 | / | 0.21 |
| 17139 | 51936.5112 | -5 | -10.1 | -50.7 | 33.3 | / | -2.61 | -9.98 | 2.24 |
| 17141 | 51938.4974 | -2.5 | -0.2 | -46.3 | 44.6 | / | -0.01 | / | -0.91 |

Table 2: Continued

| File ID | HJD -2400000 | $H_\alpha$ | | | | | Si ll 6347 | Si ll 6371 | He I 6678 |
|---|---|---|---|---|---|---|---|---|---|
| | | Wing | Abs. Cent. | V-Peak | R-Peak | Abs. Bump | | | |
| 17161 | 51959.5478 | -4.9 | 17.5 | -36.9 | 50.4 | -91.7 | 9.44 | -14.69 | -1.58 |
| 17185 | 52005.3348 | -2.6 | 6.1 | -36 | 47.6 | / | -0.48 | -24.1 | 4.7 |
| 17195 | 52027.4354 | -7.3 | -10.7 | -50.5 | 41.8 | / | -0.25 | -19.4 | -7.19 |
| 17196 | 52029.3592 | -5.6 | -9.2 | -49.4 | / | / | -9.22 | -23.87 | 6.05 |
| 17205 | 52038.5591 | -4.9 | -1.2 | -69.7 | 50.9 | / | -17.01 | -19.4 | -2.03 |
| 17003 | 52322.6262 | -7.1 | / | -45.2 | 60.3 | -113.2 | 9.57 | 1.27 | 4.25 |
| 17004 | 52322.6737 | -4.9 | / | -46.5 | 59 | -113.9 | 7.72 | -2.98 | 0.19 |
| 17333 | 52343.4844 | 5.17 | / | / | / | / | -0.42 | -24.51 | / |
| 17346 | 52352.5487 | -6 | / | / | / | / | -12.53 | -3.87 | -3.6 |
| 17349 | 52362.3617 | -7 | / | -43.1 | / | -108.5 | -9.69 | -14.69 | -7.42 |
| 17356 | 52366.3776 | -5.5 | / | / | / | -101.9 | 0.94 | -12.34 | 2.46 |
| 17365 | 52373.4292 | -1.5 | 4.9 | -35.8 | 32.3 | / | 4.3 | -2.87 | -0.62 |
| 17397 | 52417.4381 | -6.8 | / | -39.3 | 44.3 | -124.7 | -11.99 | -19.8 | -13.87 |
| 17533 | 52657.5959 | -7.9 | / | / | 25.5 | / | 16.01 | -5.69 | -6.91 |
| 18411 | 52683.5671 | -4.6 | 7.2 | -24.8 | 29.6 | / | 8.79 | 1.13 | 6.33 |
| 18431 | 52684.4712 | -6 | 9.1 | -19.2 | / | / | 6.66 | 1.6 | 4.54 |
| 18434 | 52684.5418 | -5.6 | 4.5 | -20.2 | 21 | / | 1.94 | -2.63 | 4.99 |
| 18481 | 52687.4981 | -8.54 | / | / | / | / | -1.37 | -0.04 | -1.3 |
| 18484 | 52687.5881 | -8.08 | / | / | / | / | 4.77 | -17.45 | 0.05 |
| 18495 | 52688.4042 | -5.4 | / | -47 | / | -118.7 | 5.72 | -0.28 | 2.18 |
| 18524 | 52692.4424 | -8.9 | / | -31.3 | 55 | -103 | 6.66 | 0.9 | 5.21 |
| 18537 | 52693.4215 | -7.8 | / | -37 | / | -102.8 | 0.05 | 1.6 | 2.07 |
| 19757 | 52720.317 | -6.1 | 1.9 | -51.1 | 39.8 | / | -5.62 | -15.1 | 2.07 |
| 19793 | 52721.4777 | -4.5 | 8.3 | -47.9 | / | / | -14.59 | -14.63 | -9.6 |
| 19816 | 52722.5142 | -3 | 4.8 | -47.7 | 35 | -117.6 | -14.35 | -15.1 | -5.34 |
| 19836 | 52723.4036 | -5.3 | 8.5 | -47.3 | 34.5 | -120.4 | -10.1 | -21.92 | -13.87 |
| 19837 | 52723.4325 | -5.5 | 4.1 | -45.7 | 36.5 | -116.1 | -20.02 | -12.98 | -5.79 |
| 19864 | 52724.4549 | -5.5 | 19.2 | -42 | 31.1 | -117.4 | -12.23 | -19.8 | -10.28 |
| 19926 | 52727.4657 | -7.6 | 14.4 | -40.4 | 36.8 | -122.2 | -0.42 | -15.1 | -10.05 |
| 20412 | 52734.7316 | -6.4 | / | -38.6 | 30 | -107.2 | -2.98 | -5.14 | -4.2 |
| 21118 | 52741.9524 | -1.9 | 9.8 | -31.8 | 28 | -77.5 | 0.62 | -8.18 | -5.18 |
| 21221 | 52742.6241 | -2.6 | 8 | -33.6 | 25 | / | 2.42 | -4.63 | -3.53 |
| 21222 | 52742.6307 | -2.6 | 6.2 | -33.1 | 26.1 | / | -4.25 | 10.47 | 3.38 |
| 21421 | 52744.8203 | -3.5 | 16.5 | -25.7 | 27.6 | / | 1.06 | -17.44 | -7.55 |
| 21422 | 52744.8238 | -2.8 | 11.5 | -27.5 | 26.2 | / | 7.12 | -1.74 | -1.21 |
| 21426 | 52744.8590 | -2.2 | 13.1 | -26.2 | 27.1 | -78.4 | 8.31 | 0.05 | 1.65 |
| 22111 | 52751.7782 | -7 | / | -47.4 | 58.7 | -120.7 | 2.18 | 3.02 | 2.45 |
| 22112 | 52751.7863 | -6.9 | / | -49.1 | 62.3 | -199.4 | -2.06 | 3.62 | 11.08 |
| 22114 | 52751.7979 | -3.1 | / | -44.1 | 62.6 | / | 5.21 | 5.43 | 19.14 |
| 22115 | 52751.8014 | -5.5 | / | -47.1 | 68.4 | / | 8.84 | -8.46 | 15.11 |
| 22229 | 52752.5113 | -2.8 | / | -46.8 | 64 | -122.4 | 12.03 | 3.14 | -1.3 |
| 22429 | 52754.8454 | -5.2 | 45.2 | -38.1 | 67.5 | -102.6 | 8.38 | -2.3 | 4.46 |
| 22430 | 52754.8493 | -6.1 | 44.3 | -38.9 | 66.6 | -108.7 | 2.93 | 1.92 | 6.19 |
| 22431 | 52754.8534 | -7.1 | 43.9 | -43.5 | 67.9 | -103.8 | 0.5 | -12.57 | -0.15 |
| 22432 | 52754.8576 | -6.4 | 40 | -43.3 | 66.3 | -109.5 | -1.32 | 1.31 | 8.49 |
| 22433 | 52754.8600 | -6.6 | 40.9 | -41.1 | 65.6 | -103.9 | 2.92 | 5.54 | 2.73 |
| 22434 | 52754.8624 | -8.6 | 41.8 | -44.9 | 66.4 | -114.7 | 13.23 | 0.11 | 2.15 |
| 22435 | 52754.8644 | -6.1 | 41.4 | -40.6 | 64.9 | -110.4 | 7.77 | 9.16 | -3.04 |
| 22436 | 52754.8680 | -5.3 | 40.4 | -44.6 | 65.6 | -113.2 | 7.77 | -8.35 | 5.6 |
| 22437 | 52754.8699 | -7.1 | 39.8 | -42.3 | 64.6 | -111.5 | 9.59 | -5.33 | 12.52 |
| 22438 | 52754.8733 | -10.3 | 38.9 | -43.2 | 63.5 | -110 | 4.13 | 2.52 | 11.36 |

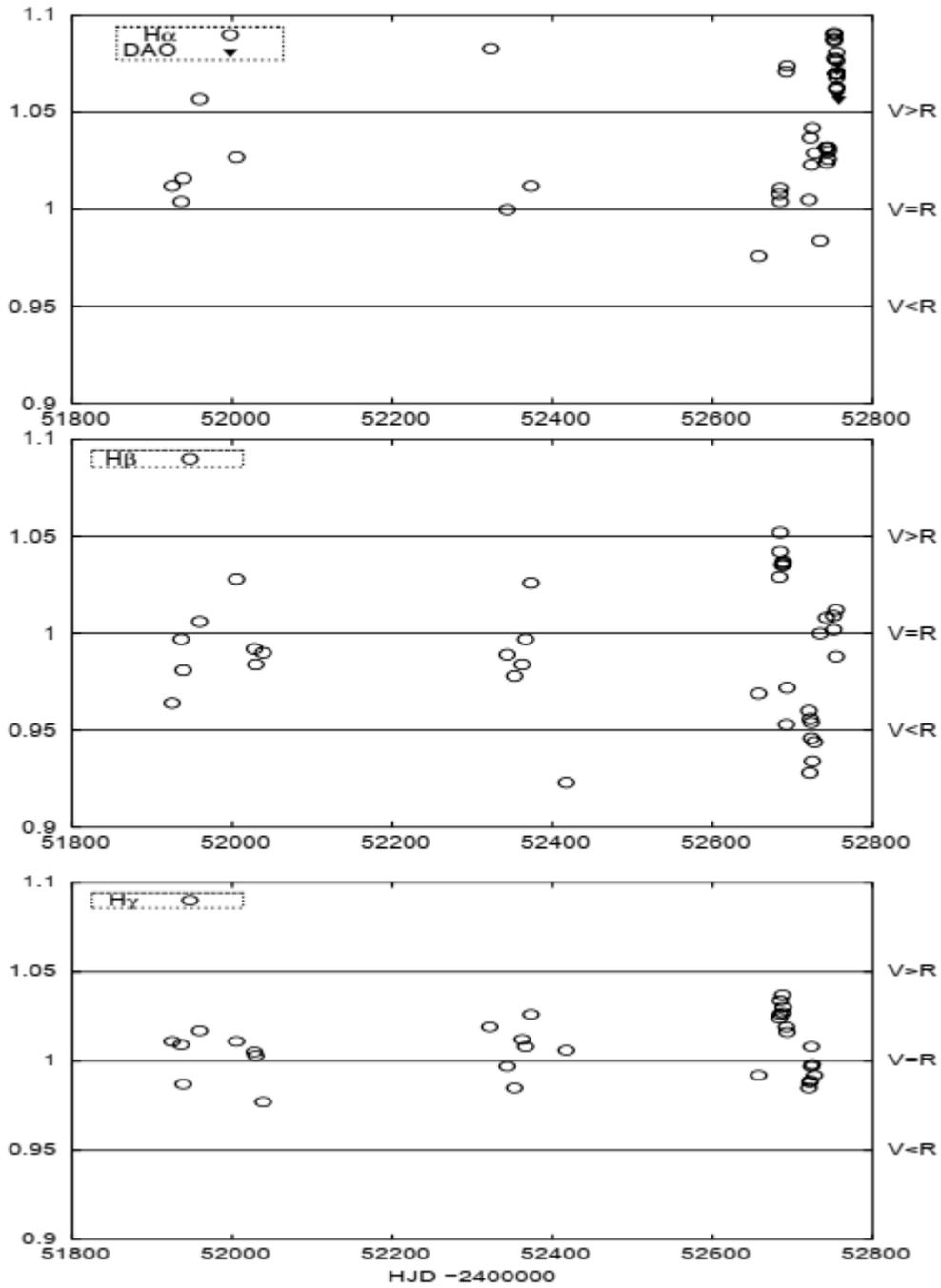

**Fig. 1.** Simultaneous time distribution of the *V/R* ratio for H*α*, H*β*, and H*γ* lines.

## 3. Period analysis

Preliminary analysis of the RV variations was important to select a suitable set of lines for further evaluation of the orbital elements. The period search was carried out separately for several line data sets (basically those which cover a long time interval of observations), where two independent numerical period searching routines were used. One is based on the phase dispersion minimization (PDM) technique (Stellingwerf 1978), the other one is a computer program PERIOD04 (Breger 1990). For PERIOD, starting values for the frequencies need to be determined by some other technique such as Fourier (which is available in the program itself) or using PDM. Within limits given by the alias structure of the observations, PERIOD can improve the frequency values by minimizing the residuals of a sinusoidal fit to the data. A search for periodicity in the interval 1-160 d has been performed for three sets of RVs, namely RVs of the $H_\alpha$ emission wings, of the metallic lines Si II 6347, 6371 Å, and of the He I 6678 Å line. The analysis confirmed the $61^d.5549$ period for two sets of lines: He I 6678 Å and Si II 6347 Å (which Juza et al. 1991) reported for the orbital period) with amplitudes of 5.9 km/s and 8.9 km/s, respectively. For the $H_\alpha$ data set, we detected two frequencies at $0.9987 cd^{-1}$ and $0.0162 cd^{-1}$ (the orbital one).

The upper and middle panels of Fig.2 represent a periodogram (frequency vs. power) of the original data, the right insets in each panel illustrate the spectral window function around the detected frequency, while the left insets represent the residuals pre-whitened for the detected frequency for both the He II 6678 Å and Si II 6347 Å lines, respectively.

In Fig. 2, the lower panel is similar to the upper and middle ones, but for the $H_\alpha$ data set, the pre-whitened spectra for a higher frequency of $0.9987 cd^{-1}$ show the disappearance of the frequencies and that of the orbital period as well. Fig.3 displays the corresponding phase diagrams for both data sets. We adopted a similar procedure for searching the period of the V/R ratio of $H_\alpha$. The frequency of $0.0162 cd^{-1}$ (P=$61^d.64$) was detected for the V/R variations (which is close to the orbital frequency), the corresponding phase variations for this frequency are illustrated in Fig.4. Phase diagrams of the V/R variations for $H_\alpha$, $H_\beta$, and $H_\gamma$ lines display consistent behavior with each other when folded with $P = 61^d.64$. Fig. 5 represents the phase diagrams folded with P = $61^d.5549$ for the Hα emission wings (upper panels) and Hα absorption center (lower panels).

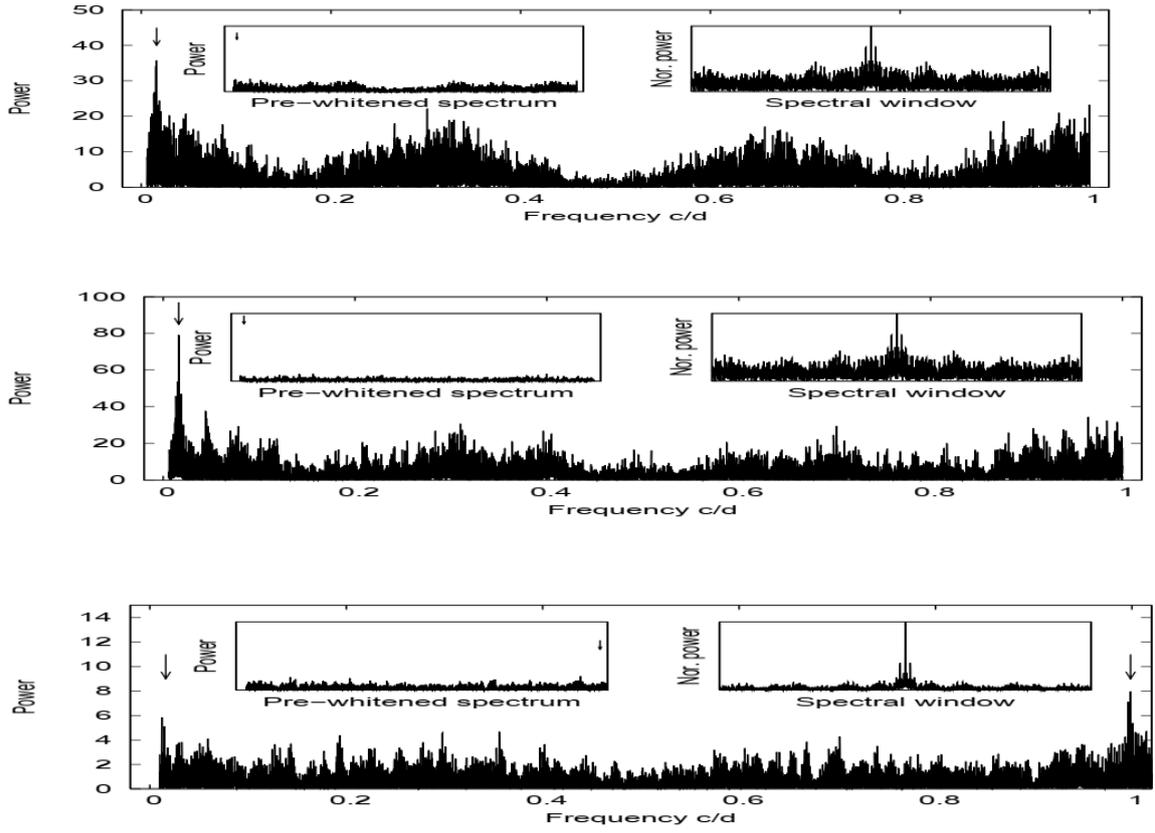

**Fig. 2.** Each main panel displays the periodogram of the original data, the other left and right panel represent the spectral window for the detected frequency and the residual after prewhitening with the detected frequency respectively. The arrows refer to the peak of the predicted frequency. The panels represent period search results for ***Upper panel:*** He I 6678A, ˚ ***Middle panel:*** Si II 6347 A and ˚ ***Lower panel:*** H$\alpha$ line.

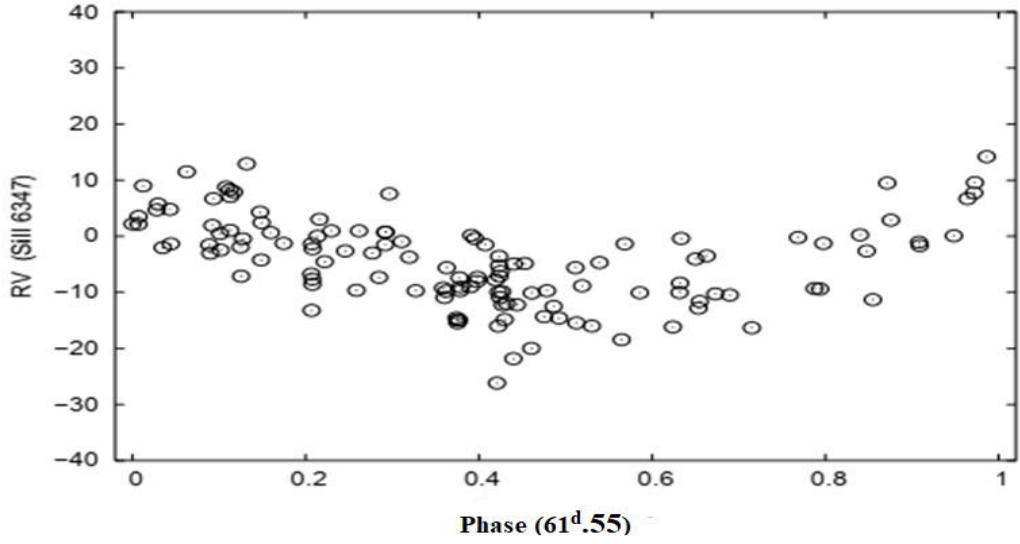

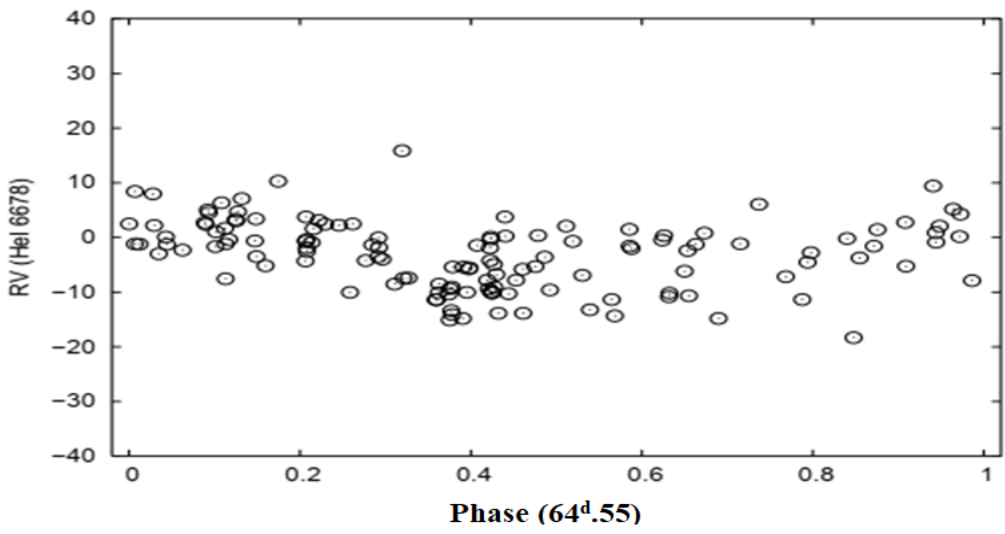

**Fig. 3.** Corresponding phase diagrams for RVs of Si II 6347 A and He ˚I 6678 A lines folded with the period ˚ 61d. 5549.

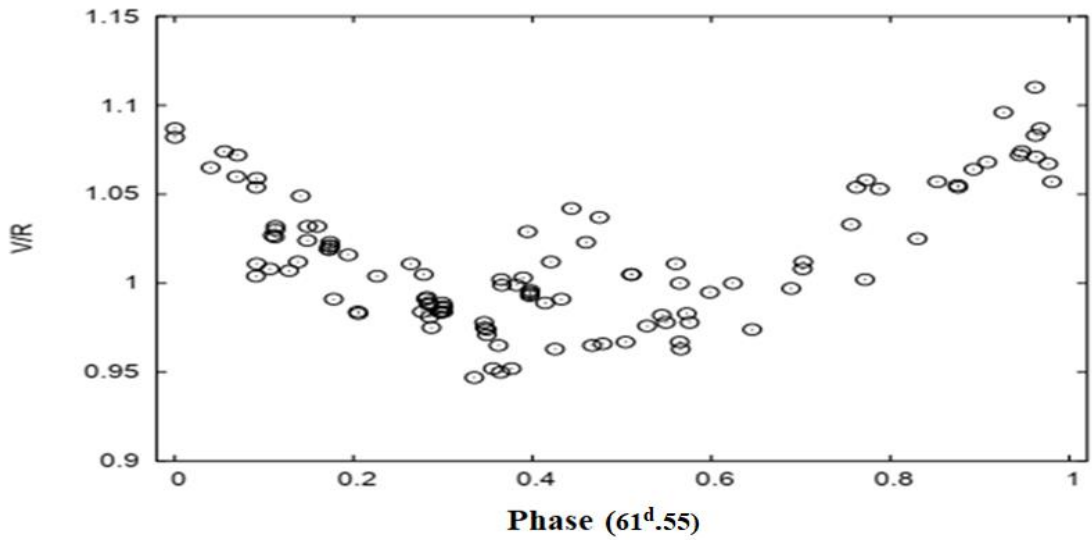

**Fig. 4.** H$\alpha$ V/R variations folded with the $61^d.55$ period.

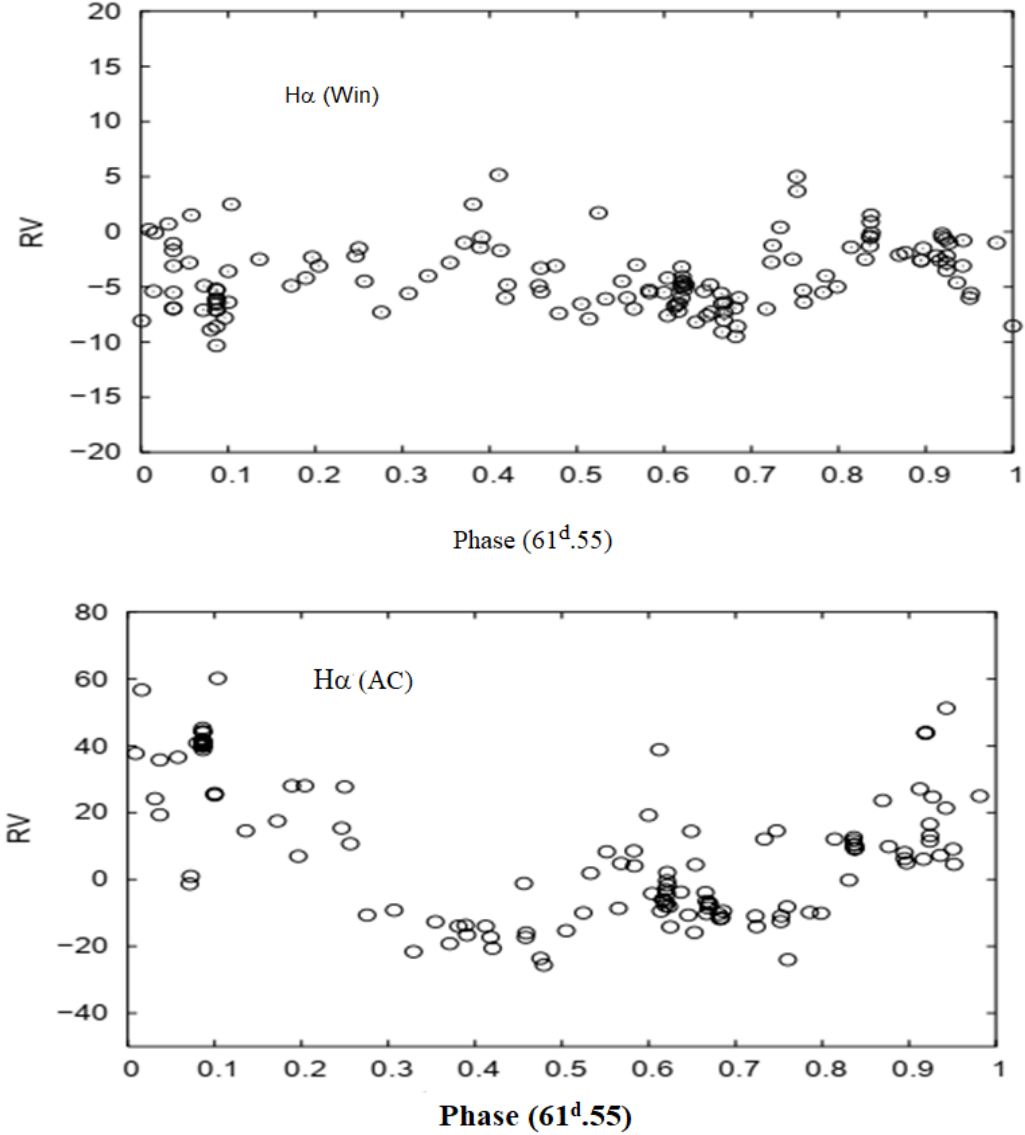

**Fig. 5.** Phase diagrams folded for ($P = 61^d.5549$) *Upper panel:* Hα emission wings, *Lower panel:* Hα absorption center.

## 4. RVs analysis and orbital elements

Orbital elements were computed from the RVs obtained by the method described in Section 2.1 and using the codes FOTEL (Hadrava 1990) and SPEL (developed by the late Dr. Horn ). FOTEL is a FORTRAN code for separate or simultaneous solving of light curves, radial-velocity curves, visual (interferometric) measurements, and eclipse timing of binary and/or triple stellar systems.

Given the results of the period analysis (Section 3), we excluded the data set of the $H_\alpha$ from the final solution, since it was dominated by a higher frequency of $0.9987 cd^{-1}$.

The values of the eccentricity **e** and the periastron longitude ω have been examined through different runs of both codes for RVs in different lines. The test of Lucy and Sweeney (1971) built-into the code SPEL indicated that the orbit is circular and small values of the eccentricity are spurious. We thus fixed e=0 and ω =0 and allowed the code to converge for the period P, the time of the periastron passage $T_{peri}$, and the RV semiamplitude K only.

The RV-curves measured for eight different lines (in addition to $H_\alpha$) have been solved to check for the constancy of the solution. Five formal solutions were obtained and given in Table 3. The first three (I to III) were obtained individually for the Si II 6347 Å, Si II 6371 Å, and He II 6678 Å lines. Let us note that solution II (obtained for Si II 6371 Å) shows a higher $rms$ than the other solutions. The line Si II 6371 Å is very weak in the spectrum of κ Dra and, in addition, it is blended with an iron line, and therefore it has a complicated structure. As a result, the phase distribution of its RVs has a larger scatter. The solution IV was obtained from the previous sets in addition to the three sets of RVs measurements in the Balmer lines ($H_\beta$, $H_\gamma$, and $H_\delta$). FOTEL also allows us to derive individual systemic RV $\gamma$ for each data set. Finally, we performed another solution V using all data sets involved in solutions IV and three other regions He I 4471 Å + Mg II 4481 Å (as one region), O I 7772-5 Å and $H_7$. We fixed the period at the value $P = 61^d.5549$ obtained by an alternative KOREL solution (cf. Section 7) and solved with the following ephemeris (similar to Juza et al. 1991)

$$T_{\max.RV} = (HJD 2449980.22 \pm 0.59) + 61^d.5549 E$$

Solution V has the smallest scatter, the smallest errors of the orbital elements, and agrees with other solutions I and III within their errors. It is thus accepted as the final solution.

Table 3: FOTEL orbital solutions for k Dra

| Element | Solution I SiII 6371 | Solution II SiII 6371 | Solution III HeI 6678 | Solution IV (I-III)Hβ,Hγ,Hδ | Solution V (I-IV) MgII,HeI OI 7772, H7 |
|---|---|---|---|---|---|
| P[d] | **61.52±0.03** | **61.60±0.06** | **61.54±0.06** | **61.59±0.03** | 61.55±0.02 |
| Tperiast. | **49979.95±1** | **49977.51±1.7** | **49980.05±1.6** | **49978.95±0.90** | 49980.22±0.59 |
| K[kms$^{-1}$] | **8.10±0.38** | **6.79±0.68** | **5.17±0.82** | **6.84±0.25** | 6.81±0.24 |
| γ(mean) | **-4.5** | **-8.7** | **-2.8** | -- | -- |
| γ(SiII6347)[kms$^{-1}$] | -- | -- | -- | -4.21 | -4.11 |
| γ(SiII6371)[kms$^{-1}$] | -- | -- | -- | -8.55 | -8.51 |
| γ (HeI6678)[kms$^{-1}$] | -- | -- | -- | -2.90 | -2.84 |
| γ (Hβ)[kms$^{-1}$] | -- | -- | -- | -2.50 | -2.47 |

| γ (Hγ)[kms$^{-1}$] | -- | -- | -- | -0.32 | -0.31 |
|---|---|---|---|---|---|
| γ (Hδ)[kms$^{-1}$] | -- | -- | -- | 0.41 | 0.39 |
| γ (MgII+HeI)[kms$^{-1}$] | -- | -- | -- | -- | 4.26 |
| γ (OI 7772)[kms$^{-1}$] | -- | -- | -- | -- | -0.81 |
| γ (H7)[kms$^{-1}$] | -- | -- | -- | -- | -0.30 |
| *f*(m)(M*sun*) | 0.0033 | 0.002 | 0.0009 | 0.002 | 0.002 |
| *Asini*(R*sun*) | 75.54 | 63.40 | 48.27 | 64.05 | 63.49 |
| **No. of RVs** | 151 | 147 | 159 | 325 | 354 |
| **rms [kms$^{-1}$]** | 5.55 | 7.92 | 5.57 | 4.83 | 4.75 |

## 5. Orbital and phase-locked variations

## 5.1 RVs variations

RVs of the red and violet emission peaks show long-term variations in terms of years. Fig.6a displays its time-distribution. These long-term variations were found for the strength and intensity of the whole line (more details in Saad et al. 2004). A signature of the $61^d.5549$ orbital period has been detected after pre-whitening it for the long-term variations period. Figs.6b and 6c illustrate the periodograms of their RVs after pre-whitening for long-term variations, with that (8044 d) cycle obtained by Saad et al. (2004) from brightness, line intensity, and equivalent width variations. The arrow indicates the peaks of the detected frequencies, which are very close to that of the orbital period. Therefore we can say that the violet and red peaks in their displacement follow two different movements of the systems, one of them is related to the circumstellar matter surrounding the system (which is the dominant one as it is clear from a time-distribution diagram), while the other expresses the orbital motion of the binary system. In this respect, we have to note that we failed to find similar behavior in other parameters with dominant long-term variations (intensity of red and violet peaks, and the line strength).

## 5.2 V/R variations

Violet-to-red peak intensity ratio (V/R) variations are one of the most striking features of the emission lines, which give a measure of the line asymmetry. Usually, the V/R ratio behaves quasi-periodic on a timescale of years. Be stars may change from a stable V/R=1 ratio to V/R variability and back (Porter & Rivinius 2003). Hanuschik et al. (1996) supposed that the majority of Be stars have emission lines with stable and equal V and R peaks, although one-third of them show a cyclic variation of V/R.

On average, periods of V/R variations are usually about seven years (Mennickent & Vogt 1991) and connected with long-term variations. In some particular cases, they display different successive phases of activity, as reported 59 Cyg (Harmanec et al. 2002) or ζ Tau (Hubert et al 1982). For several other stars, V/R variations are quasi-periodic on a long-timescale, e.g., γ Cas with a period of 9150 d, and κ CMa with a period of 8052 d (Mennickent & Vogt 1991). We found erratic or quasi-periodic variations, as reported also for β CMi (Mennickent & Vogt 1991). In most cases with V/R variations, long-term variations are found to be superimposed on short-term ones, which is sometimes misinterpreted as irregular variations. In a particular case of 59 Cyg, two significant periods have been obtained. A short one, 28.1971 d, which follows the orbital period of the binary system, and a long one of 722 d, which is connected with the formation of the new Be envelope (for more details see Harmanec et al. 2002).

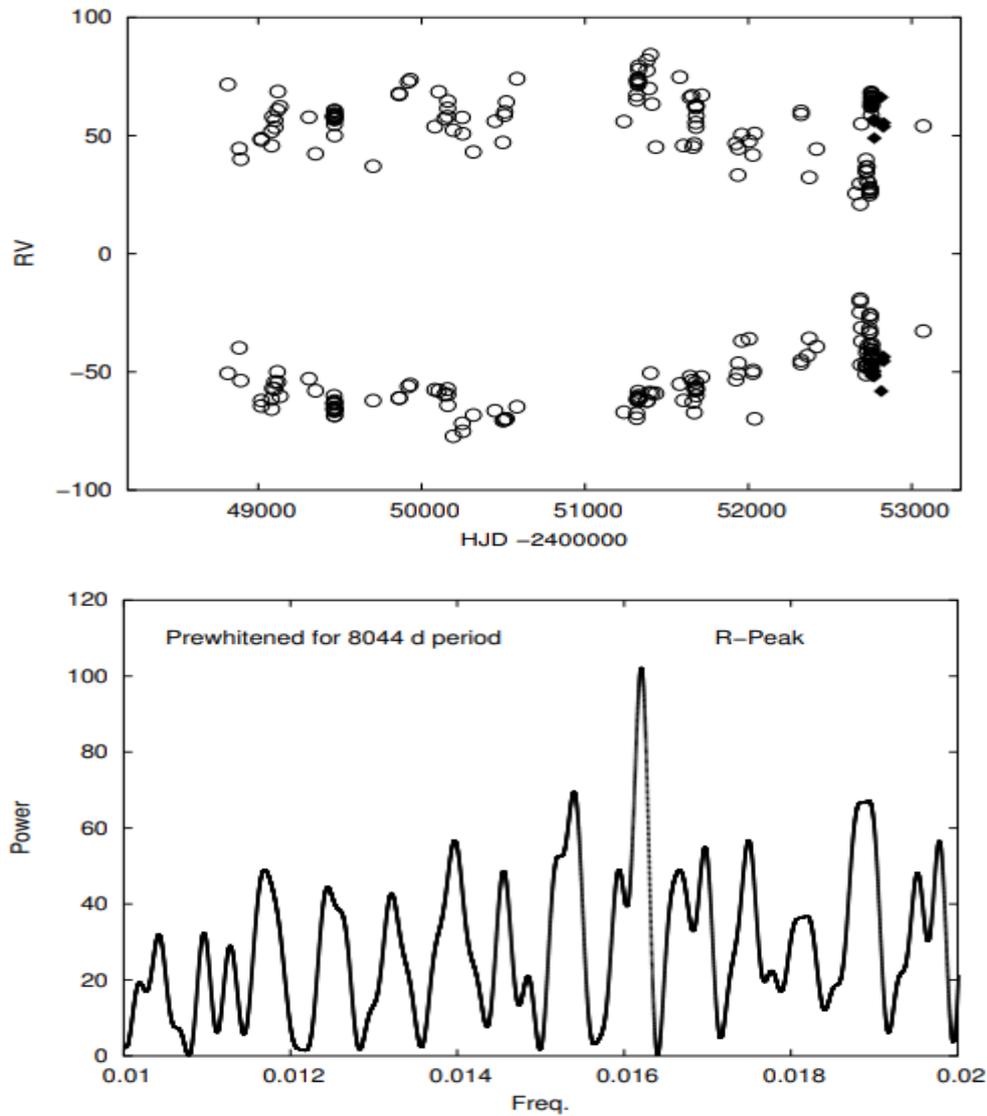

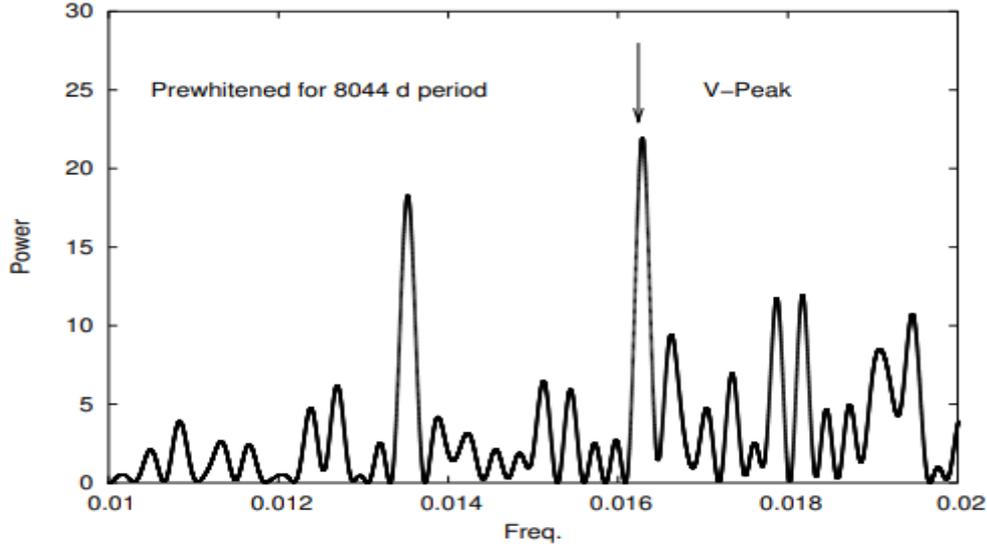

**Fig. 6.** Violet and red peaks variations of Hα. *(a):* the long-term variations of the RVs of the *V* and *R* peaks, *(b):* the periodogram of the RVs residuals of the *V* peak after prewhitening for long term variations, *(c):* similar to *(b)* but for the *R* peak. The arrow indicates the peaks of the predicted frequencies.

κ Dra is one of a few Be stars, which are known by the synchronization of the V/R variations with that of the orbital motion and behave independently of its long-term variations. Other similar systems include 4 Her, where V/R variations follow its orbital period of $46^d.1921$ period (Koubský et al. 1997), 88 Her, which varies with a $86^d.7221$ period (Doazan et al.1985), and $\phi$ Per with a period of $126^d.696$ (Poeckert 1981) are known as V/R phase-locked with their orbital period. Concerning the $H_\alpha$, the V/R asymmetry of κ Dra, Arsenijevic et al. (1994) noted that, according to the polarization results, the internal layers of the envelope have axial symmetry, but the outer $H_\alpha$ emitting region, affected by the presence of the companion, is probably associated with nonaxisymmetric external layers. In different panels of Fig.7 we give a series of simultaneous profiles of $H_\alpha, H_\beta$, and $H_\gamma$ showing variations of the V/R variations at different phases. Various postulations and models have been devoted to explain long-term V/R variations. Struve (1931) suggested an elliptical-ring model (which is a geometrical one), which was elaborated by McLaughlin (1961). Further modifications were done by Huang (1973, 1975), who attributed the V/R variations to the apsidal motion of an elliptical ring, in which emitting atoms revolve around the star according to the Kepler's law. A binary model proposed by Kříž & Harmanec (1975) is another view, where the companion star deforms the disk into the elliptical one through tidal interactions.

Kato (1983) and Okazaki (1991) explained the long-term V/R variations by one-armed global oscillations, where a thin non-self gravitational Keplerian disk is distorted by a density wave. Let us apply this model to $\kappa$ Dra. The emission structure originates in the envelope, while the binary star is sitting somewhere inside. Therefore, it affects the innermost parts of the surrounding structure. This part of the disk rotates most rapidly. Due to the Roche lobe geometry, there is a high asymmetry in the outer parts of the disk, which therefore reflects an asymmetry between the red and blue components (i.e V/R variations), alternating around the orbit. In $\kappa$ Dra, binarity seems to play the most important role in the line asymmetry (Panoglou et al., 2018), which explains the V/R phase-locked variations with the $P = 61^{\rm d}.5549$ period.

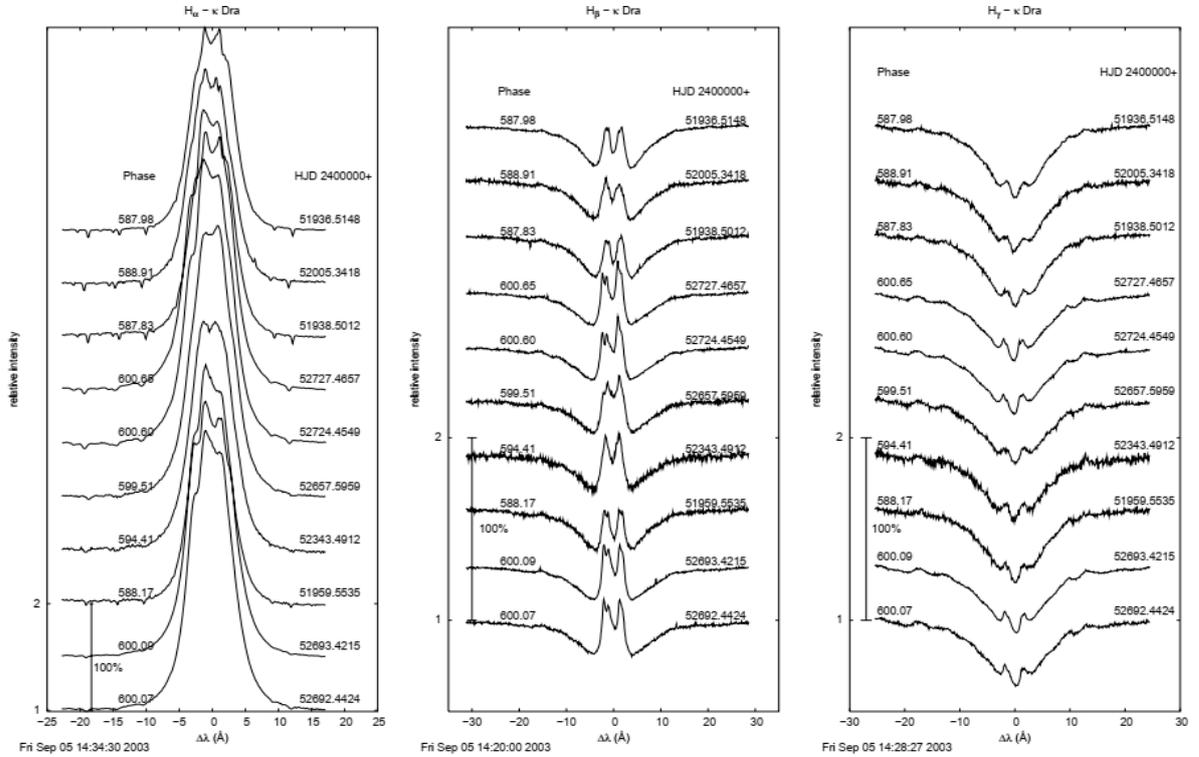

**Fig. 7.** Variations of the double-peaked profiles for different phases during one period. *Left panel:* for H$\alpha$, *Middle panel:* for H$\beta$, and *Right panel:* for H$\gamma$.

## 6. Moving absorption bumps

A traveling sub-feature from the blue to the red part of the profiles of some absorption lines on the time scale of hours has been reported by Hill et al. (1991) for $\kappa$ Dra. This feature has been already detected in the spectra of some Be stars. For 4 Her, it was discovered by Koubský et al. (1997). These authors detected this feature in the violet wing of the $H_\alpha$ line and folded it with the corresponding orbital period.

For $\kappa$ Dra Moving Absorption Bumps (MAB) were found simultaneously in the violet peaks of both $H_\alpha$ and $H_\beta$ over their observing run. They are moving through the violet peak redward, and their RV varies from -130 to -80 for $H_\alpha$ and from -100 to -20 for $H_\beta$. Their strength is variable from a deep absorption to a faint one. The feature completely disappears in some particular phases.

Its velocity-time distribution does not show long-term variations comparable to those found in brightness, equivalent width, and line intensity variations. We also found no sign of short-term variability of the MAB. In Fig.8 we illustrate this feature in $H_\alpha$ and $H_\beta$ with different intensity and velocities at the same epoch.

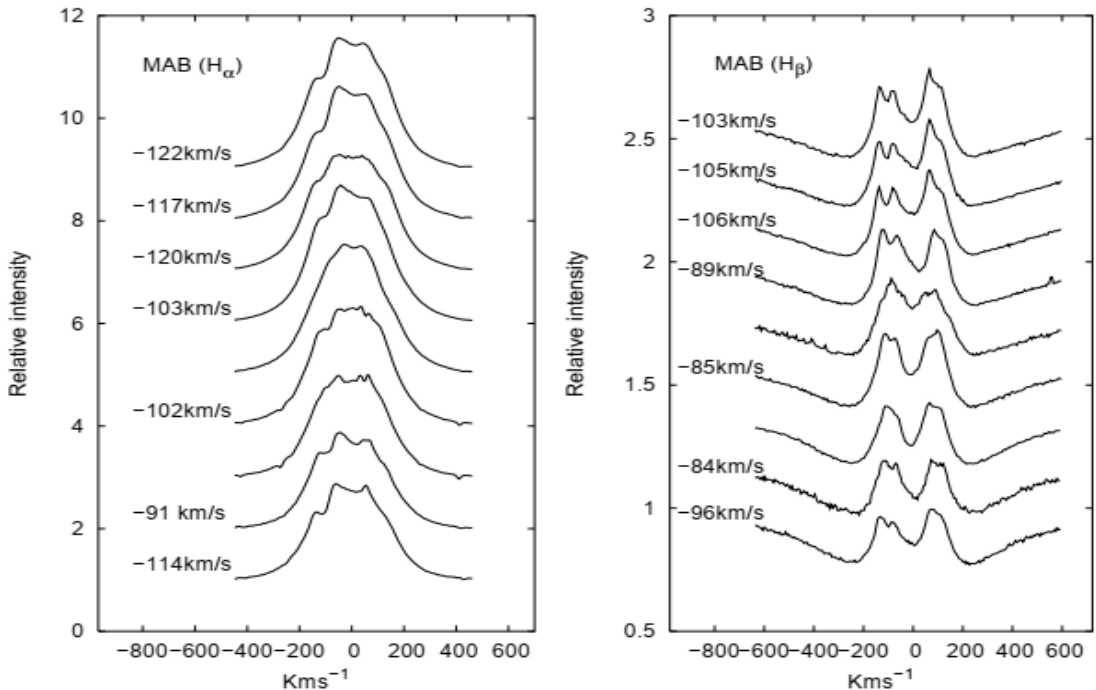

**Fig. 8.** Simultaneous plots of the absorption bump in the H*α* and H*β* lines (from Kubát & Saad 2008, Fig. 1).

With a notable scatter its variability is more probably related to the orbital period, Fig.9 shows the phase diagram of the MAB velocities folded with the orbital period for $H_\alpha$ and $H_\beta$. Although these absorption features have been a subject of many studies, no one clear theory can account for this variation. The new result found here is the simultaneous appearance in both $H_\alpha$ and $H_\beta$. Taking into account the fact that the characteristic time variation of this MAB is close to that of the orbital variation, its nature may be due to the binary interaction.

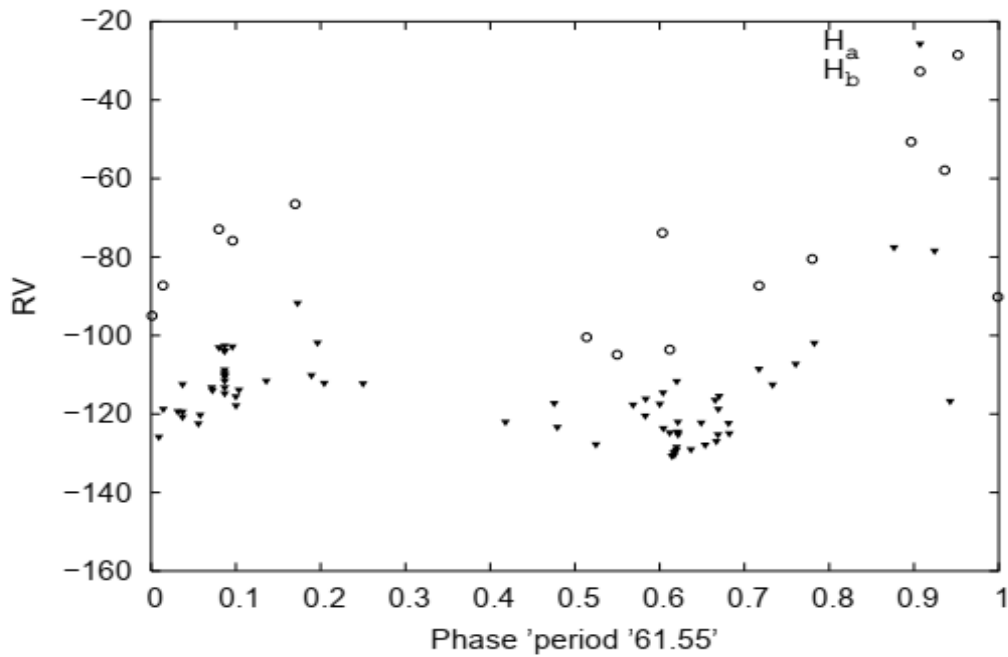

**Fig. 9.** Phase diagram of the RVs of the absorption feature folded with the orbital period for H*α* and H*β* *(see also Kubát & Saad 2008, Figure 2).*

## 7. Spectral disentangling

In addition to the RV analysis, we used the method of spectral disentangling for a more detailed analysis of the *κ* Dra spectra. The application of this method is not straightforward for this star and it thus needs a special discussion. In its original form (cf. Simon and Sturm, 1994) disentangling requires that the line-profiles of the component stars are not subject to any intrinsic variability, and all variations observed in the spectra are solely due to blending of the Doppler-shifted lines from the stellar components. The long-term variations in the strengths of the $H_\alpha$ line together with the V/R variations are the most obvious in the case of *κ* Dra. We thus took the assumptions of the code KOREL for Fourier disentangling (i.e. simultaneous decomposition of component spectra and solution of orbital elements) and 'line-photometry' of binary and multiple stars with variable line strengths (Hadrava, 1997) model to fit the observed profiles.

As a first step, we ran KOREL for one-component (plus telluric lines when necessary) solutions with variable line-strengths in different spectral regions (see Table 4). For each region, we obtained a decomposed profile. Figure 10 illustrates the comparison of the disentangled spectra of selected photospheric lines with that of the theoretical model for the following parameters ($T_{eff}$ =14000 K, log g = 3.5, and v sin i = 170 km/s), determined for $\kappa$ Dra in Saad et al. (2004).

We used KOREL to search for and disentangle photospheric lines of the secondary component, but these attempts failed for absorption lines as well as for weak emission lines even when we used a set of spectra with a reasonably high S/N ratio and carefully normalized them to increase the ability of KOREL for detection of the secondary. In all regions, no direct evidence of the secondary spectrum was found.

Table 4. KOREL solutions for k Dra (P=61.5549 d, e=0 and ω=0 are fixed)

| Region | HJD Tperi. | K [kms$^{-1}$] |
|---|---|---|
| Hα | **49963.6107** | 1.20 |
| HeI 6678 | **49980.6272** | 5.83 |
| SiII 6347 + 6371 | **49983.0273** | 4.90 |
| OI 7772-5 A | **49982.7098** | & 6.66 |
| Hβ | **49987.3446** | 4.20 |
| Hγ | **49984.4458** | & 5.13 |
| HeI 4471+ MgII 4481 | **49983.3012** | 5.09 |
| Hδ | **49976.7465** | & 7.14 |
| SiII + HeI+ OI | 49982.1262 | 5.08 |

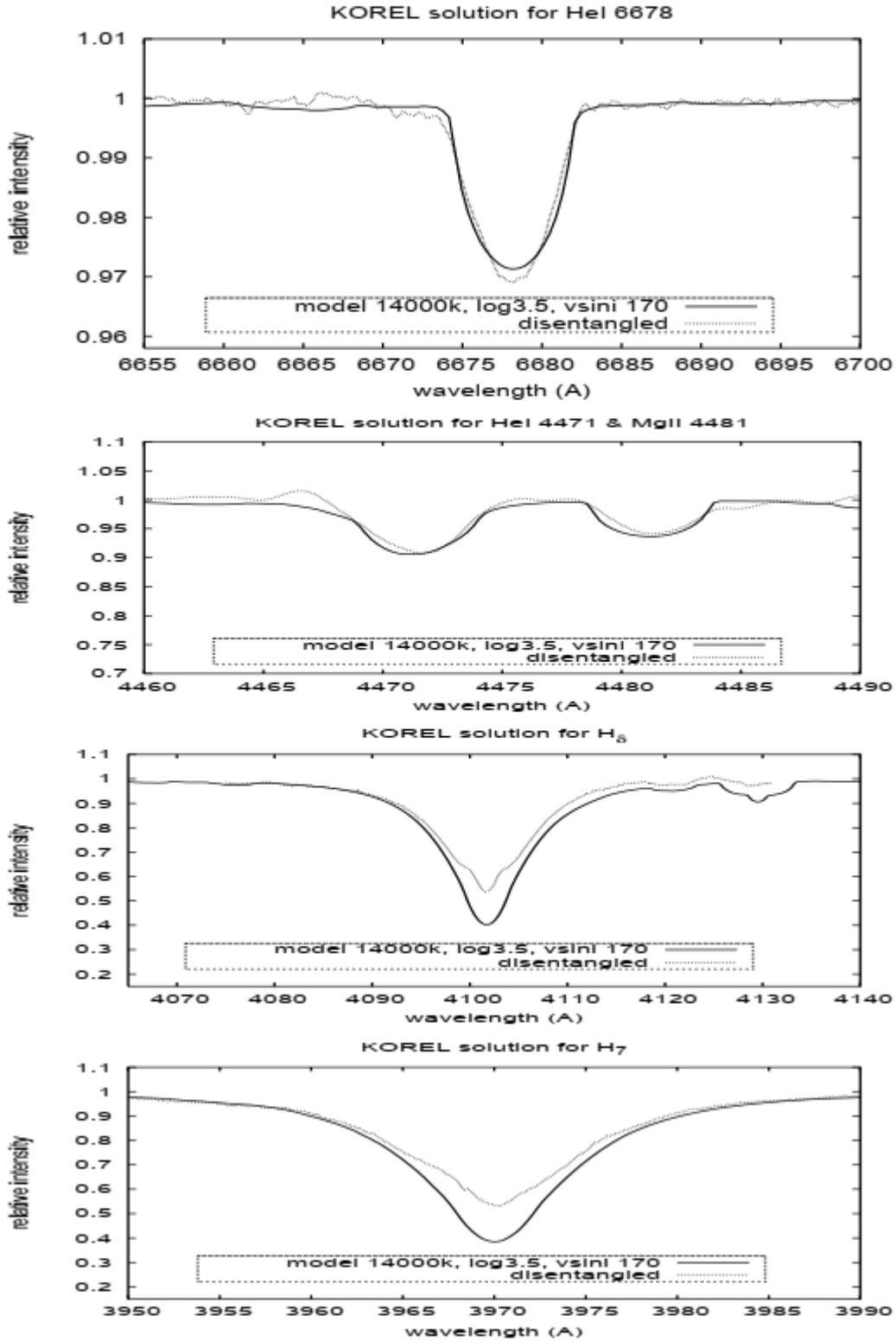

**Fig. 10.** Comparison of the disentangled spectra of some photospheric lines (dashed line) with those of a theoretical model (solid line) for parameters $T_{\rm eff} = 14\,000$ K, $\log g = 3.5$, and $v \sin i = 170$ km s$^{-1}$.

## 8. Rapid variability and line profile variations

Rapid variability and line profile variations ($lpv$) are common features in early-type stars. Be stars show various features with different scales of variations from star to star, which increase the mystery of the Be phenomenon since we can not generally accept one model for all stars.

## 8.1. Rapid variability

For $\kappa$ Dra, rapid variability was reported by Hubert-Delplace & Hubert (1979) and Andrillat & Fehrenbach (1982). Three decades ago $lpv$ of $\kappa$ Dra was studied by Hill et al. (1991). They detected rapid variations (on a timescale of hours) with a period of 0.545 days in the width and asymmetry of the absorption lines He I 4471 Å, Mg II 4481 Å, and He I 4388 Å and attributed these variations to nonradial pulsations period. However, Juza et al. (1991) showed that the detected periods of rapid variations are all aliases of one physical period and that they are related to the rotation of the star.

In this respect, we derived more constrained limits of the rotational period of the primary ($0^d.95$-$1^d.91$) in comparison to ($0^d.5 - 2^d.0$) reported by Juza et al. (1991), based on the revised values of the Be star radius 6.4 ±0.5 $R_\odot$ and projected rotational velocity v sin i = 170 km/s (Saad et al. 2004) for different inclination angles from 30° to 90°, and using explicitly the relation

$$P_{rot} = 50.633(R/R_\odot)(v\sin i)^{-1}\sin i$$

For the orbital inclination i=30°, the rotational period is 0.953 ± 0.07 days. The error accounts only for the uncertainty in the determined radius. An ultra-rapid variation on a time scale ~2min in the $H_\alpha$ profile structure detected on 14/15 Feb. 1993 (from 36 co-added and averaged profiles) is reported by Anandarao et al. (1993). The $lpv$ manifest themselves mainly through the line profile asymmetry, line width, and sometimes give rise to the moving (absorption or emission) sub-features.

For different regions of the spectra, our RV measurements were examined for rapid variability. We searched periodicity using the two above mentioned independent numerical period searching routines. As mentioned earlier (in sec. 3 ) for $H_\alpha$, the periodogram for the RV measurements of the outermost parts of the line (wings), the lower panel of Fig.2 gives a higher frequency at $0.9987 cd^{-1}$, while the other frequency is centered at the orbital period.

After the pre-whitening process (for $0.9987 cd^{-1}$) other frequencies peaks completely disappear which supports the suggestion that this rapid variation in the RV ( around one day ) is not a real one, and it is probably associated with the orbital frequency. For some absorption lines (mainly for He I 6678 Å and Si II 6347Å ) we searched for short period variations. After removing the orbital variation from the data (pre-whitened) of RV measurements of the mentioned two lines, the residuals were searched for short term variations. Two different short periods with lower powers are found for each set of data, $0^d.171198$ (at amplitude 3.775 km/s) and $0^d.416805$ (at amplitude 2.9956 km/s) for HeI 6678 Å and Si II 6347 Å respectively.

The various sets of RVs available from all lines (366 points which are previously used to obtain the final solution with FOTEL) after prewhitening it for the orbital period were searched for short term variations. The detected period of $0.^d39526$ with low power (3.321 $Kms^{-1}$) may belong to the noise level. Finally, we have used the measurements of equivalent width (EW) and line central intensity ($I_c$) of $H_\alpha$ to search for the short term variability. We searched for a period in both data sets in the range between 0.1 - 2 d first using the program PERIOD. Fig 11 represents a periodogram of the original data, a spectral window, and a periodogram of pre-whitened spectra. Two main frequencies at (ordered by power): $1.00020 cd^{-1}$ and at $2.00343 cd^{-1}$ have 75% and 65% of the confidence level (as shown in the spectral window), respectively. Using the PDM technique for the same data sets and the range of periods gave the highest peak at $f = 1.00020 cd^{-1}$ similar to that obtained using PERIOD. Prewhitening the original spectra with $f = 1.00020 cd^{-1}$, removed the one at $2.00343 cd^{-1}$, however $f = 2.0025 cd^{-1}$ at lower amplitude is still found (third panel of Fig 11).

Since the obtained frequency is around 1 day, this raises the question of whether a one-day periodicity may be caused by the cadence of the observations. To solve this problem we applied the randomizing technique suggested by Eaton et al. (1995). We created the randomizing sample for our data, then we calculated the power again for the periodicity in the same frequency range, the same results were obtained. Comparable highest peaks are found in the normalized power of the original data and those obtained from a randomized data set.

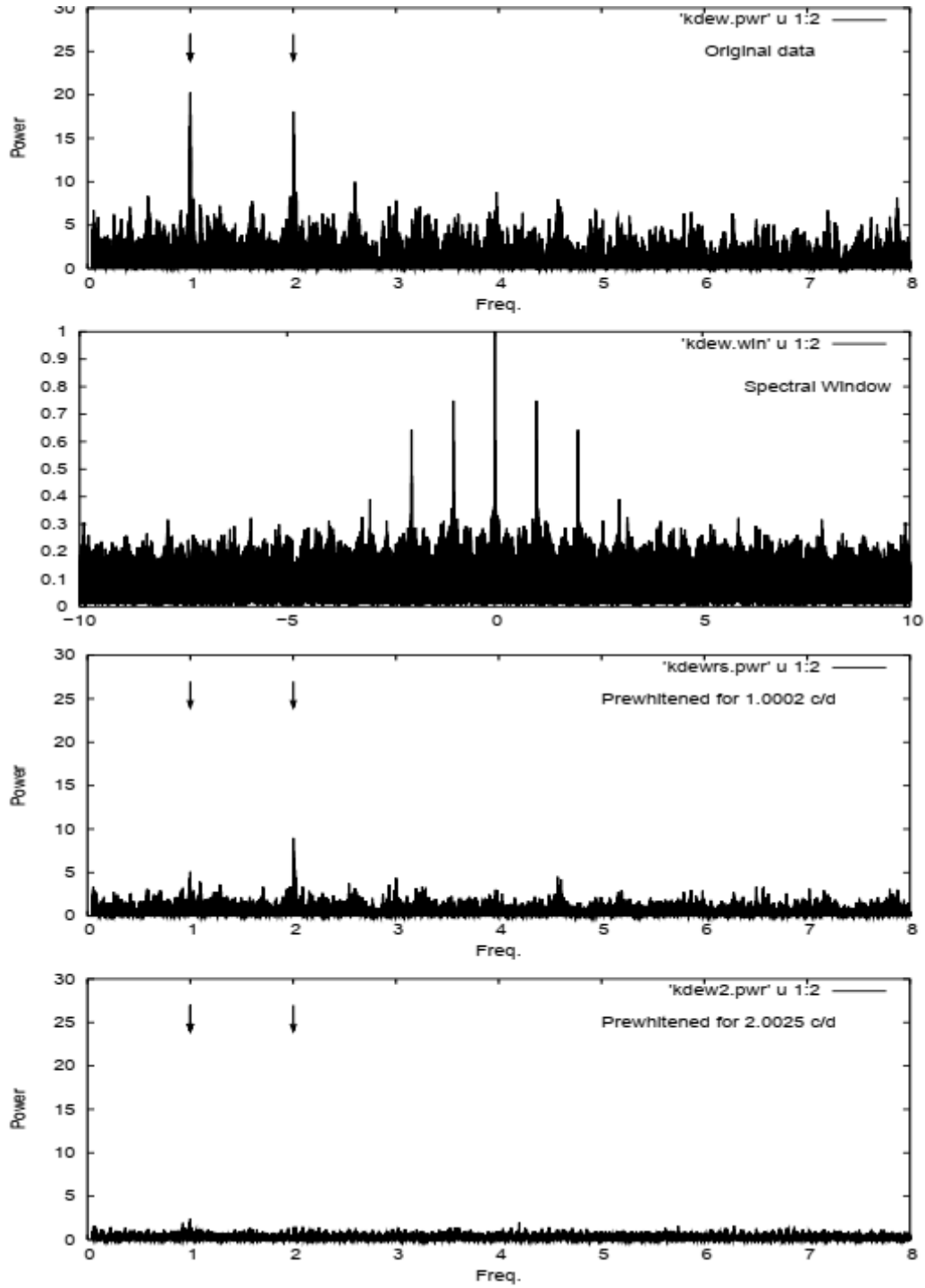

**Fig. 11.** Periodogram of the original data for the EW of H$\alpha$, spectral window, and the periodograms through the process of successive prewhitening for the detected short-term period.

## 8.2. Line Profile Variation

The profile variations (*lpv*) of the He I 6678 Å absorption line were investigated using inspection of the residuals (subtraction of the individual spectra from the average one). Two sets of closely time spaced observations of this line were obtained. Eleven spectra were obtained in the course of two successive nights, on 23/24 April 1994, between epochs 49466.3311 - 49467.5566 (~29h) using a Reticon detector in the coudé spectrograph. Another ten spectra were obtained during a 40-min period of the night of April 24 (at JD 52754.8454 - 52754.8733) using the coudé spectrograph and the CCD detector. The lower panel of Fig.12 displays the residuals of the individual spectra obtained by subtraction from the average one, and the time evolution of the corresponding profiles of individual lines, respectively. Each set of observations was related to different phases of line strength (the first set obtained in 1994 corresponds to phases of higher line strength in comparison to that obtained in 2003) as was already known from the study of long-term variations of the line. The deviation from the mean profile is clear in some spectra. The upper panel of Fig.12 shows the line wings affected by the emission. The residuals of the individual spectra from the averaged one show fluctuations from one exposure to the other, however, no clear sequence of variation can be detected. No night-to-night changes similar to that found in the Be star FX Lib (as reported by Guo, 1994) were found in the $H_\alpha$ line profile of $\kappa$ Dra.

**Fig. 12.** *Upper panel:* Line profiles of He I 6678 A observed on 24 April 2003 and 23/24 April 1994. Each profile is fitted with the average one (dashed lines). The numbers on the right hand side mean the midexposure in HJD. *Lower panel:* The residuals of the individual spectra are subtracted from the average one.

## 9. Conclusions

We reanalyzed the spectroscopic data of the bright Be star $\kappa$ Dra obtained from June 1992 to April 2003. We used the generally accepted orbital parameters derived with two codes FOTEL and KOREL. It suggested a circular orbit with a period of $61^d.5549$ and semi-amplitude of RV variations of the primary component of the binary system (K= 6.81 $Kms^{-1}$). Based on the FOTEL solution, the estimated values of the projected separation of the system components $A \sim 126.9$ R$_\odot$ and the mass function $f(m) = 0.002 M_\odot$ were derived.

V/R variations were measured for $H_\alpha, H_\beta, H_\delta$, and some other photospheric lines. They are found to be phase-locked with the orbital motion. Rapid variability is investigated and for different line parameters, it is related to the rotational velocity of the star. No night-to-night variations were found in the line profiles of $\kappa$ Dra. The presence of an absorption bump traveling in phase with high negative RVs may be explained by some cold clumps in a distant outflowing gas. It is noted that from the KOREL disentangling of different regions of the spectra, no direct evidence of the secondary spectrum was found for Balmer and other metallic lines in both optical and near-infrared spectra.

Taking into account the spectroscopic mass of κ Dra ($M_1$ = 4.8 $M_\odot$) determined by Saad et al. (2004) and the results of the FOTEL solution, we calculated several possible values of the secondary mass $M_2$ for different inclination angles. If the rotation axis of the primary and the orbital axis are parallel, then for the stellar radius R = 6.4 $R_\odot$ and rotation velocity 170 km s$^{-1}$ (Saad et al., 2004) the minimum inclination angle i ≈ 27°. Thus if κ Dra rotates near its break-up velocity, the most appropriate inclination angle is about 30°. A corresponding mass of the secondary is then $M_2$ ≈ 0.8 $M_\odot$.

**Acknowledgments.** The research is supported by the National Research Institute of Astronomy and Geophysics (NRIAG) and the Science and Technology Development Fund (STDF No. 5217). This paper used spectroscopic data from the archive of the Perek 2-meter Telescope and has made use of NASA's Astrophysics Data System Abstract Service. The criticism of the first version of this article by the anonymous referee and the revision of the text done by Dr. J. Kubát, greatly helped us to improve the article and the arguments, and the authors gratefully appreciate this effort. S. Saad would like to thank Prof. Dr. P. Hadrava, (the author of the KOREL and FOTEL codes) for his helpful discussions when using the codes, and to express her great thanks to the system administrators at Czech Astronomical Institute, Ondřejov.